\newcommand{\squiggle}{SQuIGG$\vec{L}$E }

\newcommand{\prospector}{\texttt{Prospector}}

\documentclass[ RNAAS, 11pt]{aastex631}
\usepackage{graphicx} 
\usepackage[utf8]{inputenc}
\usepackage{amssymb}

\usepackage{tabularx}
\usepackage{booktabs}
\usepackage{amsmath}
\usepackage [english]{babel}
\usepackage [autostyle, english = american]{}

\begin{document}

\title{Meet the Neighbors: Gas Rich ``Buddy Galaxies'' are Common Around Recently Quenched Massive Galaxies in the \squiggle Survey}

\author[0009-0005-4226-0964]{Anika Kumar} 
\affiliation{Laboratory for Multiwavelength Astrophysics, School of Physics and Astronomy, Rochester Institute of Technology, 84 Lomb Memorial Drive, Rochester, NY 14623, USA} 
\affiliation{Department of Physics and Astronomy and PITT PACC, University of Pittsburgh, Pittsburgh, PA 15260, USA}

\author[0000-0003-4075-7393]{David J. Setton}\thanks{Brinson Prize Fellow}
\affiliation{Department of Astrophysical Sciences, Princeton University, 4 Ivy Lane, Princeton, NJ 08544, USA}

\author[0000-0001-5063-8254]{Rachel Bezanson}
\affiliation{Department of Physics and Astronomy and PITT PACC, University of Pittsburgh, Pittsburgh, PA 15260, USA}

\author[0000-0001-9820-9619]{Alan Pearl} 
\affiliation{Argonne National Laboratory 9700 S Cass Ave Lemont, IL 60439}

\author{Erin Stumbaugh} 
\affiliation{Department of Physics and Astronomy and PITT PACC, University of Pittsburgh, Pittsburgh, PA 15260, USA}

\author[0000-0003-3256-5615]{Justin~S.~Spilker}
\affiliation{Department of Physics and Astronomy and George P. and Cynthia Woods Mitchell Institute for Fundamental Physics and Astronomy, Texas A\&M University, 4242 TAMU, College Station, TX 77843-4242, US}

\author[0000-0002-1759-6205]{Vincenzo~R.~D'Onofrio}
\affiliation{Department of Physics and Astronomy and George P. and Cynthia Woods Mitchell Institute for Fundamental Physics and Astronomy, Texas A\&M University, 4242 TAMU, College Station, TX 77843-4242, US}

\author[0000-0002-5612-3427]{Jenny E. Greene}
\affiliation{Department of Astrophysical Sciences, Princeton University, 4 Ivy Lane, Princeton, NJ 08544, USA}

\author[0000-0002-1714-1905]{Katherine A. Suess}
\affiliation{Department for Astrophysical \& Planetary Science, University of Colorado, Boulder, CO 80309, USA}

\author[0000-0003-1535-4277]{Margaret E. Verrico}
\affiliation{University of Illinois Urbana-Champaign Department of Astronomy, University of Illinois, 1002 W. Green St., Urbana, IL 61801, USA}
\affiliation{Center for AstroPhysical Surveys, National Center for Supercomputing Applications, 1205 West Clark Street, Urbana, IL 61801, USA}

\begin{abstract}

In this work, we characterize the environments of massive ($\log(M_\odot/M_\star)\sim11.2$) $z\sim0.7$ post-starburst galaxies (PSBs) by studying serendipitously-detected CO(2--1) emitters found in targeted observations of the \squiggle sample. We report $31\pm6\%$ of the galaxies from this survey host nearby gas-rich ``buddies'' with stellar masses $\geq 10^{10},M_\odot$ and molecular gas comparable to their central PSBs ($M_{H_{2}} \sim 10^{10} M_\odot$), but $\sim0.8$ dex lower stellar mass ($\sim 10^{10.4} M_\odot$). Based on their location in position-velocity space, each buddy is consistent with being bound to the haloes of their \squiggle host galaxies. We compare to the UniverseMachine model and find that \squiggle galaxies host a typical number of neighbors for their stellar mass, suggesting that PSBs live in environments typical of co-eval similarly-massive galaxies.

\end{abstract}

\section{1. Introduction}
Galaxy color and morphology bimodality implies a process--known as ``quenching"--wherein blue star forming disks transform their morphologies, exhaust their gas supply, and emerge as red-and-dead ellipticals. However, the mechanisms driving this transition remain poorly understood. Previous works have established two quenching modes: a slow process dominating at low-z and the fast process at high-z \citep[e.g.,][]{galaxyzoo}. To understand the rapid formation of the most massive galaxies at early times \citep[e.g.,][]{Thomas2005}, we must identify the physical processes driving fast quenching. Post-starburst galaxies (PSBs) experience a sharp decline in star formation within the past $\leq1$~Gyr, and are therefore the direct products of fast-track quenching \citep[e.g.,][]{dressler_gunn}. Although rare locally, intermediate-redshift studies offer a unique opportunity to build large samples of PSBs, near enough to facilitate detailed, multi-wavelength studies \citep[e.g.,][]{Setton2023}.

Observational and theoretical studies highlight the role of mergers in triggering rapid quenching \citep{maggie_2023,hopkins2008}. Furthermore, early quenched galaxies appear to live in high-z overdensities \citep[e.g.,][]{degraff2024}. Therefore, it is natural to consider whether massive PSBs at much later times reside in similar, but late-forming, overdensities. The opposite is found locally ($z<0.3$), where PSBs tend to live in less dense regions than older quiescent galaxies \citep[e.g.,][]{Yesuf}. Less is known about the environments of PSBs at cosmic noon, in part due to the challenge of systematically obtaining spectroscopic redshifts. Because neighboring satellite galaxies are often gas-rich, spectroscopic environmental studies can be performed using CO lines \citep[e.g.,][]{phibbs}. In this work, we utilize ALMA CO observations of 51 massive PSBs at $z\sim0.7$ to characterize their environments using their fortuitously detected gas-rich neighbors (known as `buddy galaxies').

\section{2. Data \& Methods}

The \squiggle Survey (Studying QUenching in Intermediate-redshift Galaxies: Gas, angu$\vec{L}$ar Momentum, and Evolution Survey) is a multi-wavelength study of PSBs at $z\sim0.7$. A detailed analysis of the stellar populations and spectroscopic identification can be found in \cite{suess2022}. The survey now spans 51 galaxies mapped in CO(2--1) \citep{setton2025}. Here we perform a systematic search and analysis of these gas-rich neighbors in the full dataset. By visually inspecting all CO(2--1) cubes and HSC i-band images \citep{Aihara_2022}, we flag candidate CO detections that spatially coincided with HSC detections. We then fit the 1D CO spectra extracted in a $1''$ aperture with Gaussians of free amplitude, velocity, and dispersion using \texttt{SciPy curve$\_$fit} (see Figure 1). We detect neighbors around 16 \squiggle galaxies at $\ge 3\sigma$.

We utilize photometry from DECaLS-DR9 \citep{decal_survey} and HSC-PDR3 \citep{Aihara_2022} catalogs to fit spectral energy distributions (SED) of our sample. When HSC-PDR3 coverage is available, we adopt its \(g,r,i,z,y\) bands supplemented with \emph{WISE} \(W1\) and \(W2\) measurements from DECaLS. In regions not covered by HSC, we utilize DECaLS \(g,r,z,W1,W2\) data. Only measurements with $S/N\geq\,5\sigma$ are included in the fits. We pair \prospector~\citep{johnson:17} with \texttt{dynesty}~\citep{dynesty} nested sampler to fit the SEDs. Our setup closely follows \cite{suess2022} (see Section~3.1), except we adopt a parametric delayed-\(\tau\) model to derive the star formation histories.

\section{3. Results}

Figure 1B compares the PSB targets and buddy stellar masses. All buddies are systematically less massive ($\sim0.8$~dex) than their PSB counterparts. Because the buddies are less massive, we next test whether they are gravitationally bound to the PSBs by examining their positions in velocity–radius phase space (Figure 1C). Given the median stellar mass of the central PSBs ($\log(M_\odot/M_\star)\sim11.2$), we assume a density profile corresponding to a dark matter halo of $M\sim10^{13}M_{\odot}$ and derive velocity dispersions following \cite{lokas2001, prada2012}. These estimates are conservative, as halo masses are likely more than a hundred times greater than stellar masses. Figure 1C shows the relative velocity versus radial distance from each host galaxy, normalized by velocity dispersion and \(r_{200}\) of a $10^{13}M_{\odot}$ halo. We over plot lines of constant $\left(\frac{r}{r_{200}}\right)\times\left(\frac{\Delta v}{\sigma_{v}}\right)=0.1$. All buddies lie within the virialized region, and therefore are considered satellites.

To evaluate whether this corresponds to an over-density of satellites, we compare to the expected radial distribution of galaxies around similarly massive galaxies $(\geq \log(M_\odot/M_\star)\sim11.2)$ in the $z\sim0.75$ snapshot of UniverseMachine Data Release 1 \citep{behroozi2019}. To compute theoretical comparisons, we calculate the number of massive ($M_{*}\geq\,10^{10}\,M_{\odot}$) satellites within $\Delta\,V\leq\pm500\mathrm{km/s}$, centered on host galaxies above $2\times10^{11}\,M_{\odot}$. Figure 1D shows the observed cumulative number of galaxies as a function of distance (teal band) along with our theoretical comparison (gray line). We note that our measurement is only a lower limit; we only identify buddies with detectable molecular gas ($\log(M_{H_2})\gtrsim10$), whereas we can only apply a stellar mass cut to the UniverseMachine model. Nonetheless, we find good agreement between our observed and predicted distributions of satellite galaxies. This result suggests that the high detection rates of satellite galaxies around PSBs can be explained by their high masses and do not necessarily imply that they live in significant overdensities.  

\begin{figure} [t] 
\centering
\includegraphics[scale=2,width=0.5\textwidth]{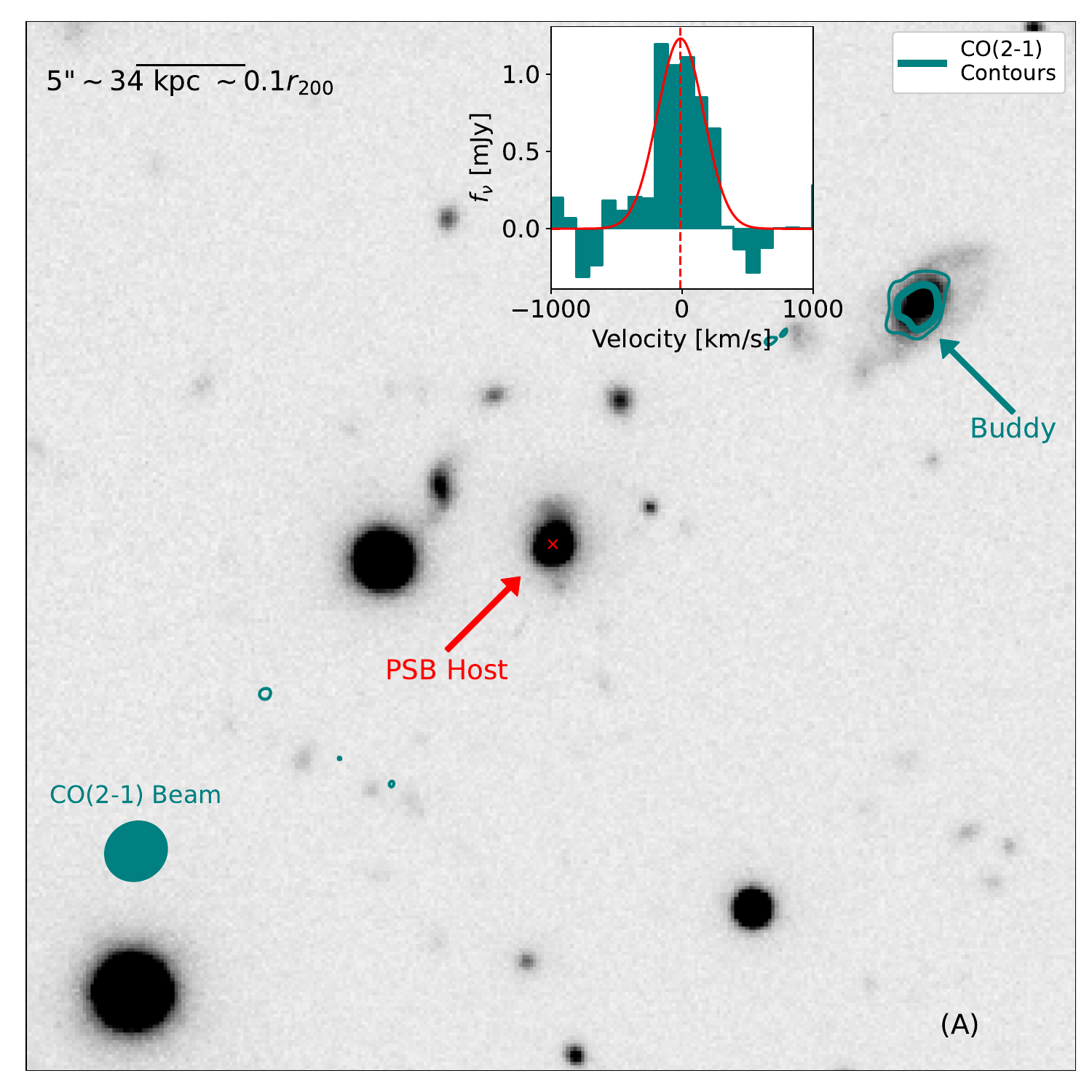}
\includegraphics[scale=2,width=1.0\textwidth]{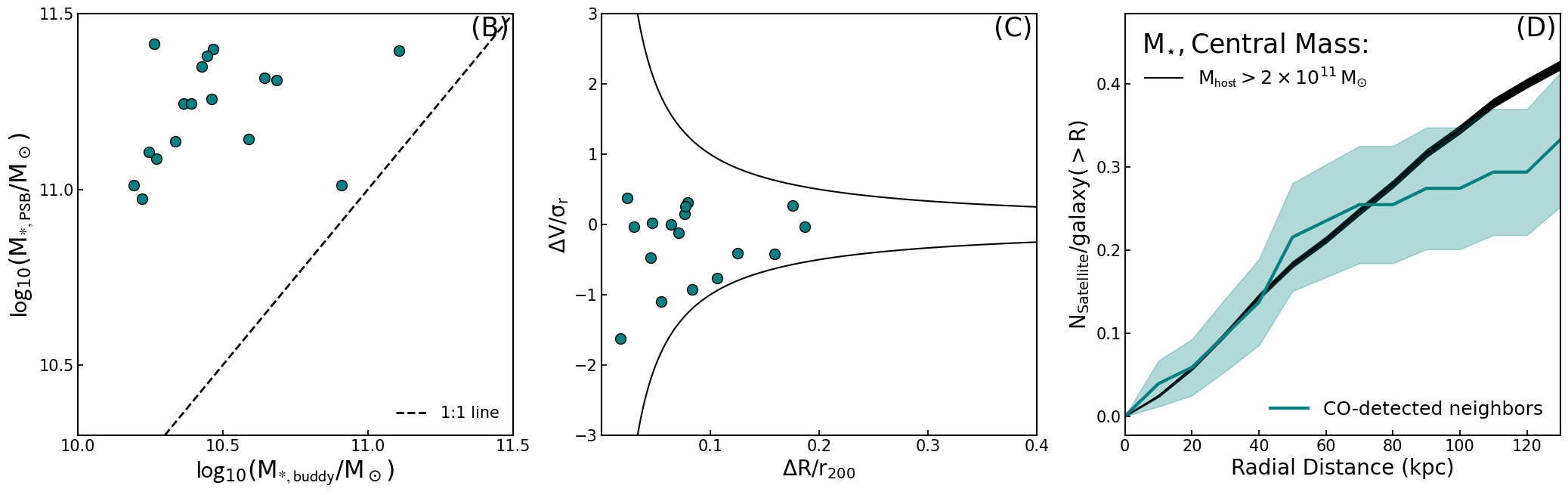}
\caption{Panel A: A 48''$\times$48'' HSC-i image centered on an example \squiggle galaxy, J1017-0003. Although the primary target (red x) is undetected in CO(2--1), it hosts a CO(2--1)-detected ``buddy'' (teal 3 and 5$\sigma$ contours). Also shown is the CO(2--1) spectrum measured in a $1''$ aperture and the Gaussian best fit (red). The best fitting velocity (relative to the \squiggle host) is shown as a dashed line. Panel B: Comparison of central and satellite masses, Panel C: shows the satellites plotted in phase space, along with lines at $\left(\frac{r}{r_{200}}\right)\times\left(\frac{\Delta v}{\sigma_{v}}\right)=0.1$ to show the virialized region. Panel D: the radial distribution of satellites around \squiggle galaxies, with a comparison to galaxies of similar mass from the UniverseMachine.} \label{noteplot}

\end{figure}


\section{Acknowledgments}
 The authors gratefully acknowledge support from NSF-CAREER grant AST-2144314, NSF-AAG grant AST-1907697, and computational resources from Research Computing Services at the Rochester Institute of Technology \citep{https://doi.org/10.34788/0s3g-qd15}. For data related acknowledgments, see \cite{setton2025}.


\bibliography{bib}
\end{document}